# Titan's Wind and Unanticipated Temperature Asymmetry

T. Kostiuk[1,8], T. A. Livengood[2,8], T. Hewagama[3], G. Sonnabend[1,4,8], K. E. Fast[1], K. Murakawa[5,8], A. T. Tokunaga[6,8], J. Annen[1,8], D. Buhl[1], F. Schmülling[7]

---

[1] *NASA Goddard Space Flight Center, Code 693, Greenbelt, Maryland 20771*

[2] *Challenger Center for Space Science Education, 1250 North Pitt Street, Alexandria, Virginia 22314, USA.*

[3] *University of Maryland, Department of Astronomy, College Park, Maryland 20742, USA.*

[4] *National Research Council Associate at NASA Goddard Space Flight Center*

[5] *National Astronomical Observatory of Japan, Subaru Telescope, Hilo, Hawaii. Present address ASTRON, P.O. Box 2, 7990 AA, Dwingeloo, The Netherlands*

[6] *University of Hawaii, Institute for Astronomy, 2680 Woodlawn Drive, Honolulu, HI 96822*

[7] *I. Physikalisches Institut, Universität zu Köln, Zülpicher Str. 77, 50937 Köln, Germany*

[8] *Visiting astronomer at Subaru, which is operated by the National Astronomical Observatory of Japan*

**Saturn's largest satellite, Titan, has stratospheric wind speeds that may be up to ~210 m/sec[1], circling Titan in about a day compared to Titan's slow 16-day rotation. Theoretical models to explain such super-rotating winds are not well constrained[2,3,4,5,6] by observations of Titan or a similar slow rotator, Venus. Titan's stratospheric temperature and the abundance of the important constituent ethane ($C_2H_6$) should be zonally invariant due to long photochemical and radiative lifetimes and rapid transport and mixing by high zonal winds. Here, we report new measurements of zonal wind in Titan's equatorial region, including the Cassini Huygens probe entry latitude[7]. Prograde zonal winds of 190±90 m/sec are found from Doppler-shifted ethane gas infrared emission near 12 µm wavelength, confirming high wind velocity even near the equator. Surprisingly, we find a temperature difference of 6±3K between the East and West hemispheres of Titan at ~240 km altitude. Common direct processes such as diurnal heating by sunlight do not adequately explain the asymmetry, suggesting other processes, such as unknown atmospheric dynamics, may be responsible. The origin of the temperature difference is enigmatic and currently unexplained.**

The magnitude, the direction, and the global and altitude distribution of Titan's winds and temperature remain important parameters for understanding the atmospheric dynamics and meteorology of slowly rotating bodies[3,4] and for interpretation of data retrieved from *Cassini* and its *Huygens* probe, particularly from the *Huygens* Doppler Wind Experiment[8,9]. The wind velocity in Titan's stratosphere was first measured directly using infrared heterodyne spectroscopy (IRHS)[10] of ethane emission lines formed in the ~0.6 mbar pressure region of Titan's atmosphere[1], ~210 km altitude, yielding a mean prograde wind 210±150 m/s and a global ethane mole fraction[11] of $(8.8\pm2.2)\times10^{-6}$. Other techniques have been applied to investigate Titan's winds as well, including indirect determination from thermal emission maps[12], and stellar occultation[2,13,14], which do not establish flow direction. Direct tracking of cloud features by ground-based or Hubble Space Telescope imaging has been prevented by a lack of suitable cloud structures. Millimeter wave interferometry[15] of Doppler-shifted spectral lines has been successful in determining direction and magnitude of gas velocity in Titan's atmosphere, similar to IR heterodyne measurements. Retrieval of the wind direction and magnitude has been demonstrated from a measurement of Doppler shifts in solar visible spectra reflected from Titan's haze, observed with the UVES instrument at the European Southern Observatory's (ESO's) Very Large Telescope (VLT)[16]. All measurements to date have yielded global circulation velocities at various altitudes on Titan from ~50–250 m/s prograde .

New Titan observations were obtained December 18, 2003 at the 8.2 meter Subaru telescope on Mauna Kea, Hawaii, benefiting from Subaru's large aperture to obtain superior spatial resolution and signal-to-noise ratio in the retrieval of velocity information (Fig. 1) compared to earlier measurements. The Goddard Space Flight Center infrared heterodyne spectrometer, Heterodyne Instrument for Planetary Wind And Composition (HIPWAC) used for these measurements has three spectral resolution modes: 25 MHz and 5 MHz spectral resolution using a radio-frequency (RF) filter bank, and up to 1 MHz resolution with an acousto-optic spectrometer (AOS)[17]. Spectral resolving power $\geq 10^6$ fully resolves the shape of the individual ethane emission line profiles and enables highly precise determination of the absolute line frequency as well as information on temperature and constituent abundance. Ideally, absolute transition frequencies can be measured to better than $1:10^8$ and thus measure winds to ~2 m/s, as previously demonstrated for Venus[18].

Figure 1 presents measured spectra from the East and the West of Titan's disc. Net integration time on each position was 45 minutes. The field-of-view (FOV) covers an extended portion of Titan with varying viewing angle and line-of-sight velocity projection. Titan's disc is modelled as a combination of many sub-resolution elements within the FOV[19]. The emergent spectrum from each element is calculated with a radiative-transfer code, then the contribution of the individual elements is weighted and co-added to produce a synthesized spectrum. Synthesized spectra are compared with measured spectra to retrieve Doppler shift, ethane abundance, and a scale factor. The scale factor is an estimated quantity that accounts for optical losses external to the instrument. Fitting the scale factor accounts for small deviations from the estimated value due to atmospheric seeing and tracking errors. Initially, the "recommended" *Huygens* science team thermal profile[20] was used for both East and West regions on Titan. Ethane line frequencies, intensities, and $C_2H_6$–$N_2$ pressure-broadening parameters were obtained from literature[21,22,23,24]. The constant-with-height $C_2H_6$ mole fraction, the scale factor, the gas velocity and the effective topocentric velocity, are varied iteratively until the best fit to the observed spectrum is determined.

The East and West spectra were fitted simultaneously to retrieve parameters consistent with both, including the zonal wind velocity. The Gaussian probability distribution, centred at the mean wind velocity, is integrated from zero velocity to infinity to obtain the total probability of wind direction (see [1] for details). The wind velocity uncertainty is determined by fixing the other fitted parameters and varying the velocity to fit the spectrum and retrieve confidence limits on its value. Both the high- and low-resolution filter bank data retrieve a nearly identical mean horizontal wind, prograde with 98% probability and a velocity $v_{HOR}$=190±90 m/s (1$\sigma$ uncertainty). The AOS spectrum and retrievals are consistent within experimental uncertainties.

Systematic errors in the absolute frequency are minimized by alternating 5-minute measurements between East and West, and by stabilization of the $CO_2$ laser local oscillator to a drift $\leq$ 1 MHz/hr (~10 m/s)[10,18]. Random errors in tracking were <0.1 arcsec over one integration period, introducing negligible effects to the observed line frequency. Contributions from Titan's rotational velocity of 11.7 m/s were accounted for in the analysis. The altitude region probed by the observed $C_2H_6$ transitions is illustrated in Fig. 2 with the contribution functions for the principal $C_2H_6$ line and for frequencies in the line's wings, as well as the temperature-pressure profile used in the analysis (solid lines). The contribution function describes the effectiveness with which a region with particular values of temperature, pressure, and abundance combine to form the detected emission at a fixed frequency. The line wings are formed progressively deeper in the atmosphere since the lower absorption coefficient (line intensity) at those frequencies permits radiation to be transmitted through a greater gas column. The peak of the contribution (line centre) is at ~240 km (0.5 mbar) with full line contributions spanning ~130–340 km (~5–0.1 mbar).

Inspection indicates a stronger integrated emission from the eastern side of Titan's disc. This is supported by a much lower retrieved mole fraction of $3\pm1.5\times10^{-6}$ from the West spectrum when the data are fit individually, versus $9\pm5\times10^{-6}$ retrieved from the East or $8\pm3\times10^{-6}$ from the simultaneous E-W fit. While the East retrieval is consistent with prior results, the low retrieved value for the West is strongly inconsistent with all previous measurements[11,25,26]. Ethane is a stable species, with a photochemical time constant >100 years in Titan's atmosphere[27] and the high zonal winds would rapidly mix ethane horizontally. It is, therefore, unlikely that local chemistry or variability in abundance can be responsible for the East-West asymmetry.

Misalignment of the West beam could explain the lower measured intensity. There was no indication of misalignment while tracking. We tested for the possibility of a misalignment by attempting to fit the data with synthetic spectra from various positions fully or partially on Titan's disc looking for a combination where the mole fraction is the same at both positions. No combination of offset positions resulted in retrieving the same ethane mole fraction from both the East and West spectra. In addition, the spectrum modelled on the West, using the East parameters, did not reproduce the shape of the spectral line in the West spectrum (Fig. 1). This argues strongly that a pointing error is not the source of the difference between them. There is no evidence of instrumental changes in the filter bank that might result in an East-West variation in the line intensity measured, and the AOS measured a similar intensity difference. Both East and West spectra were measured the same way over nearly the same time by alternating beam positions as described earlier, thus reducing or eliminating any systematic drift in instrument properties.

A difference in temperature between the eastern and the western side of Titan in the altitude region probed can account for the difference in emission intensity. We tested this hypothesis by fixing the ethane mole fraction at $9\times10^{-6}$ and then fitting the East and West spectra by varying the temperature profile. The resultant best-fit thermal profiles for the East and the West are given in Fig. 2.

The fit to the East spectrum as expected recovers the "recommended" profile[20], while the West spectrum requires a cooler temperature. The temperature decrease at the peak of the ethane contribution function, at ~240 km is 6±3 K. There is no ready explanation for this unanticipated difference in stratospheric temperature between the two hemispheres. As a slowly rotating body, the initial expectation is that Titan would display uniform temperature in an East-West comparison. A diurnal variation due to solar heating is not readily consistent with the observed warmer dawn hemisphere in the East. The radiative time constant of about a terrestrial year also precludes radiative heating processes as a likely source of an observable effect. Atmospheric dynamical effects that have not yet been identified in Titan may be responsible for the asymmetry.

In the geometry of December 18, 2003, the geocentric central meridian longitude was 299°. Due to the projection of the FOV on Titan's disc, the East beam position covered ~310° to ~30° longitude, including the sub-Saturn point at 0° longitude. Since Titan's rotation is tidally locked to its orbit, Titan always presents the same hemisphere to Saturn. There may be a difference between the Saturn-facing hemisphere and the "anti-Saturn" hemisphere, although it is not immediately clear how this orientation might lead to a local temperature increase in the stratosphere. Analyzing in the same manner previous measurements[1], which were acquired at different Titan orbital positions, may help investigate this result further. High resolution-mode spectral imaging by the *Cassini* Composite Infrared Spectrometer (CIRS)[28] may shed more light on this apparent global temperature variation.

The retrieved prograde zonal wind on Titan, at a speed of 190±90 m/s, is high for the equatorial region and, within the uncertainties, is comparable to that obtained over more extended regions on Titan[1]. This suggests that at the probed altitudes and over the regions observed the mean winds are not significantly lower than any jet streams at high latitudes proposed by others[2,6,13]. Based on our measurements and current models such existing jets would approach supersonic speeds at the high latitudes.

Our measurements probe the 130-340 km altitude range in Titan's stratosphere in the equatorial region, including the 10.5° South entry latitude of the Cassini Huygens Probe. The lower portion of the altitude range partially overlaps the region below 160 km that will be probed by the *Huygens* Doppler Wind Experiment[8,9]. A comprehensive altitude structure of circulating winds can be established from *Cassini*, *Huygens,* and ground based measurements with HIPWAC as well as newer techniques such as UVES on ESO's VLT (probing ~100-200 km)[16] and mm-wave studies probing above ~350 km[15].

Results illustrate how limited our knowledge and understanding are of the atmosphere of Titan in particular and of atmospheric phenomena in general. However, the retrieved wind, ethane abundance and stratospheric temperature provides complementary data with temporal and global context to that obtained by the Probe as well as that obtained by the *Cassini* orbiter. Together this information will improve our understanding of the complex and unusual atmosphere of this unique satellite of Saturn and the atmospheric dynamics of slowly rotating bodies.

Correspondence and requests for materials should be addressed to T.K. at kostiuk@gsfc.nasa.gov

Acknowledgements

We thank Prof. Hiroshi Karoji, Director of the Subaru Telescope, for his support of this project. We also are grateful to Prof. Tetsuo Nishimura, Dr. Robert Calder and the Subaru support staff for their advice and technical and physical support in interfacing HIPWAC to the Subaru telescope and during our observations. A special thanks to Paul Rozmarinowski (GSFC) for his superb design and fabrication support, Juan Delgado (U. Maryland) for his invaluable and dedicated assistance in preparation of HIPWAC for this observing run, and Prof. Rudolf Schieder (University of Cologne) for the loan of the AOS. This Letter is based on data collected at the Subaru Telescope, which is operated by the National Astronomical Observatory of Japan. This work was supported by the NASA Planetary Astronomy Program.

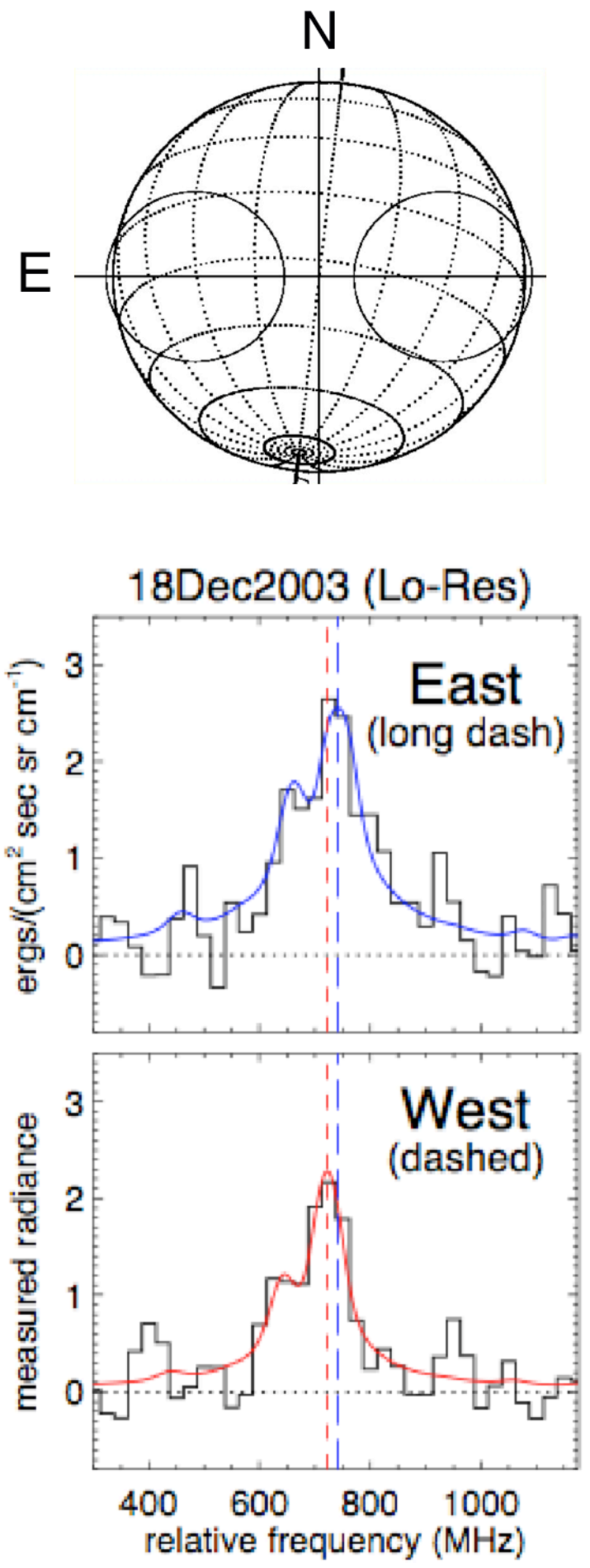

Figure 1. Observed positions on Titan and corresponding measured spectra. Top: Geometry for HIPWAC measurements on December 18, 2003, using the National Astronomical Observatory of Japan Subaru Telescope. East and North are defined as viewed on the sky. Measurements reported here were acquired following southern summer solstice in January 2003. Previous IR heterodyne measurements of Titan wind were made using NASA's 3-meter Infrared Telescope Facility (IRTF) in 1993, 1995, and 1996 near equinox[1] (November 1995). Current measurements can be compared to earlier results to investigate seasonal effects. The current 0.36 arcsec field-of-view (FOV) covers ~15° N to 50° S latitude on the 0.89 arcsec Titan disc, and overlaps the Huygens entry latitude at ~10.5°S. The FOV in the earlier IRTF measurements was ~1 arcsec, diluting the beam in limb-directed measurements. The Subaru beam geometry provides spatial discrimination to test global circulation models[2,5], which include high-velocity jet streams at high latitudes as well as expected lower-velocity flow in the equatorial region. Bottom: Spectra of Titan $C_2H_6$ emission near a central line of rest frequency 851.505 $cm^{-1}$, showing measured (histogram) and modelled (solid line) spectra from the East and West of Titan's disc. Vertical dashed lines show retrieved principal-line frequencies. The East limb spectra (blue, long dash) are at higher frequency than the West spectra (red, short dash), indicating prograde wind of 190±90 m/s. The RF filter bank at 25 MHz (0.00083 $cm^{-1}$, shown) and 5 MHz (0.00017 $cm^{-1}$) resolution and the AOS at 1 MHz resolution yield similar results. The horizontal velocity was calculated using a rigid atmospheric rotation model within the probed latitudes, using the measured frequency difference between East and West and accounting for the range of viewing angles within our FOV, and the ~25° axial tilt of Titan[1]. The slightly different latitudinal position on Titan are accurately accounted for in the analysis. Residuals at ~400 and ~950 MHz, particularly apparent in the West spectrum, approximately coincide with weak $C_2H_6$ features for which spectroscopic parameters currently are not well known. Residuals at these frequencies thus do not reliably represent the stochastic noise level.

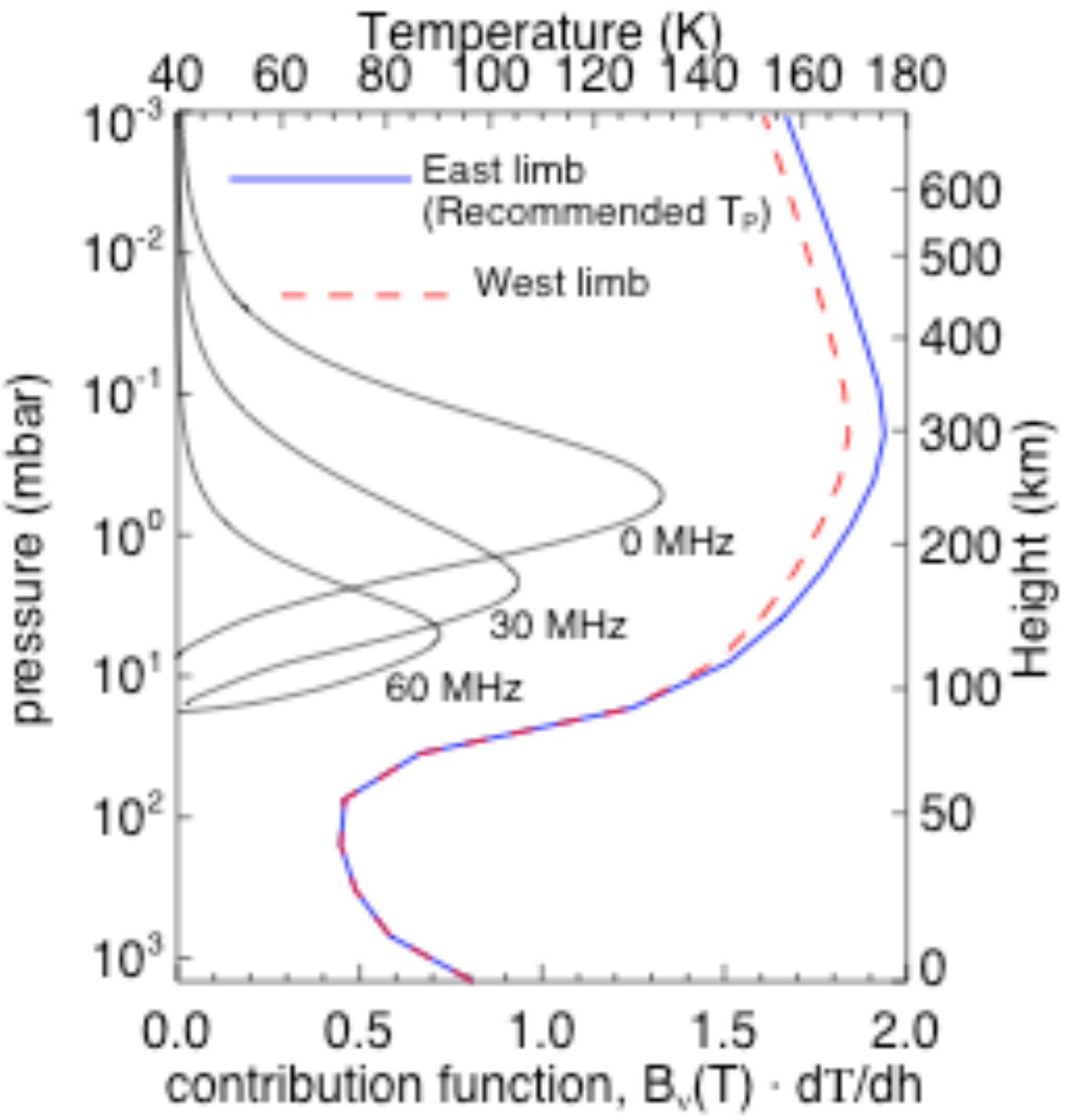

Figure 2. Contribution functions and atmospheric model. The *Huygens* “recommended” thermal profile used in the velocity retrievals and the contribution functions for the line peak and wings (solid black/blue lines). The contribution functions define the ~130–340 km (5–0.1 mbar) region probed by the primary $C_2H_6$ line, with the maximum contribution of the line peak at ~240 km altitude (0.5 mbar). A constant-with-height ethane mole fraction was used to model the spectra. Difference in E vs. W stratospheric temperature as derived for a fixed $C_2H_6$ abundance is also given (dashed line). The East temperature profile (blue) is consistent with the recommended profile. The Western profile (red) is ~6K cooler at the peak of the contribution.